\documentclass[preprint,11pt]{emulateapj}

\shorttitle{X-Shaped Radio Galaxies and the GWB}
\shortauthors{Roberts, Saripalli, \& Subrahmanyan}
\usepackage{datetime}
\usepackage{rotate}

\begin{document}

\title{THE ABUNDANCE OF X-SHAPED RADIO SOURCES: \\IMPLICATIONS FOR THE GRAVITATIONAL WAVE BACKGROUND}

\author{David H.\ Roberts,$^1$ Lakshmi Saripalli,$^2$ and Ravi Subrahmanyan$^2$}
\affil{$^1${Department of Physics MS-057, Brandeis University, Waltham, MA 02454-0911 USA} \\ $^2${Raman Research Institute, C.\ V.\ Raman Avenue, Sadashivanagar, Bangalore 560080, India}} \email{roberts@brandeis.edu}

\begin{abstract}
Coalescence of super massive black holes (SMBH's) in galaxy mergers is potentially the dominant contributor to the low frequency gravitational wave background (GWB). It was proposed by \citet{M2002} that X-shaped radio galaxies are signposts of such coalescences, and that their abundance might be used to predict the magnitude of the gravitational wave background. In \cite{PaperI} we present radio images of all 52 X-shaped radio source candidates out of the sample of 100 selected by \cite{C2007} for which archival VLA data were available. These images indicate that at most 21\% of the candidates might be genuine X-shaped radio sources that were formed by a restarting of beams in a new direction following a major merger. This suggests that fewer than 1.3\% of extended radio sources appear to  be candidates for genuine axis reorientations (``spin flips''), much smaller than the 7\% suggested by \citet{L1992}. Thus the associated gravitational wave background may be substantially smaller than previous estimates. These results can be used to normalize detailed calculations of the SMBH coalescence rate and the GWB. 

\end{abstract}


\section{INTRODUCTION}

The gravitational wave background from SMBH coalescence during galaxy mergers is proving to be a diagnostic of galaxy formation models because of the useful constraints on the GWB from improved understanding of timing noise in pulsar data \citep{Sh2013}.  Recent studies of the expected rate of merger-based SMBH formation based on current observations (corrections to BH-bulge relation, and new estimates of galaxy merger timescales) \citep{McW2013, Se2013a, Ra2015} suggest that the rates derived earlier from semi-analytic models for galaxy formation implemented in dark matter simulations \citep{Se2008, Se2010} may be seriously underestimated. This has led to the prediction of imminent detection of the GWB by pulsar timing arrays \citep{Si2013}. On the other hand, the GWB from coalescing supermassive binaries might be substantially diminished via environmental interactions \citep{Se2013b, Ra2014} and owing to the ``last parsec'' problem \citep{Mi2003, C2014, V2014}.

Axis precession has long been invoked to explain certain inversion symmetric morphologies in radio galaxies where presence of close companion galaxy and binary SMBH models have been used in some cases to explain the phenomenon \citep{E1978, Be1980,  Hu1984, T1990, Du2006, E2006, S2006, Sa2013}.  \cite{M2002} suggested that the structures of radio galaxies, which are hosted by giant ellipticals, offer independent and direct clues to the prevalence of SMBH binaries. In particular, they pointed out that ``X-shaped'' radio galaxies (XRG's) may be signposts of SMBH coalescence and hence their abundance amongst the extended radio source population might be an indicator of the true coalescence rate. Using the estimate of the fraction of radio galaxies that show X morphology from \cite{L1992} and a relic lobe lifetime of $10^8 \mbox{ yr}$ they estimated the merger rate to be about $1 \mbox{ Gyr}^{-1}$galaxy$^{-1}$.

X-shaped radio sources in particular, and the directional stability of the jets in active galactic nuclei in general, may be probes of the coalescence rate and SMBH binary formation (e.g., \cite{E1978, Be1980, G1982, D2014}). However, the relevance of XRG's in this context depends on their formation mechanism -- whether their radio structures arise from reorientations of the BH spin axes (spin flips), or slow precession or drifts in the axes, or whether the X-shaped structures are caused by the hydrodynamic deflection of backflows or jet-shell interactions or have nothing to do with changing directions of the central beams (for references and a discussion of the various proposed mechanisms see \cite{PaperI}. These different mechanisms may be operating in different cases, generating the variety in radio structures; therefore, it is crucial to determine the specific abundance of ``genuine XRG's'' that are candidates for reorientations of SMBH spin axes as these are what are relevant for the GWB.  All of the above mechanisms for generating off-axis distortions lead to radio structures with low axial ratios, and to deduce the precise mechanism in individual cases might be difficult. For this reason, good quality radio, optical, and X-ray studies of populations of complete samples of extended radio sources that have small axial ratios are critical to the problem. 

In \cite{PaperI} we presented an initial step in this direction using available data in the Very Large Array archives on the sample of 100 XRG candidates compiled by \cite{C2007} from the NRAO FIRST survey. Characterization of the extended structure in all of the sources showed that a large majority of the sample 33 of 52 sources, or 63\%) constitutes sources where the off-axis emission is traced to individual lobes. These sources  were further divided into those where the emission occurs near the inner ends of the lobes (25 of 52 sources, or 48\%) and those where the emission occurs near the outer ends (8 of 52 sources, or 15\%). 

\begin{figure}[t] 
\includegraphics[width=0.90\columnwidth]{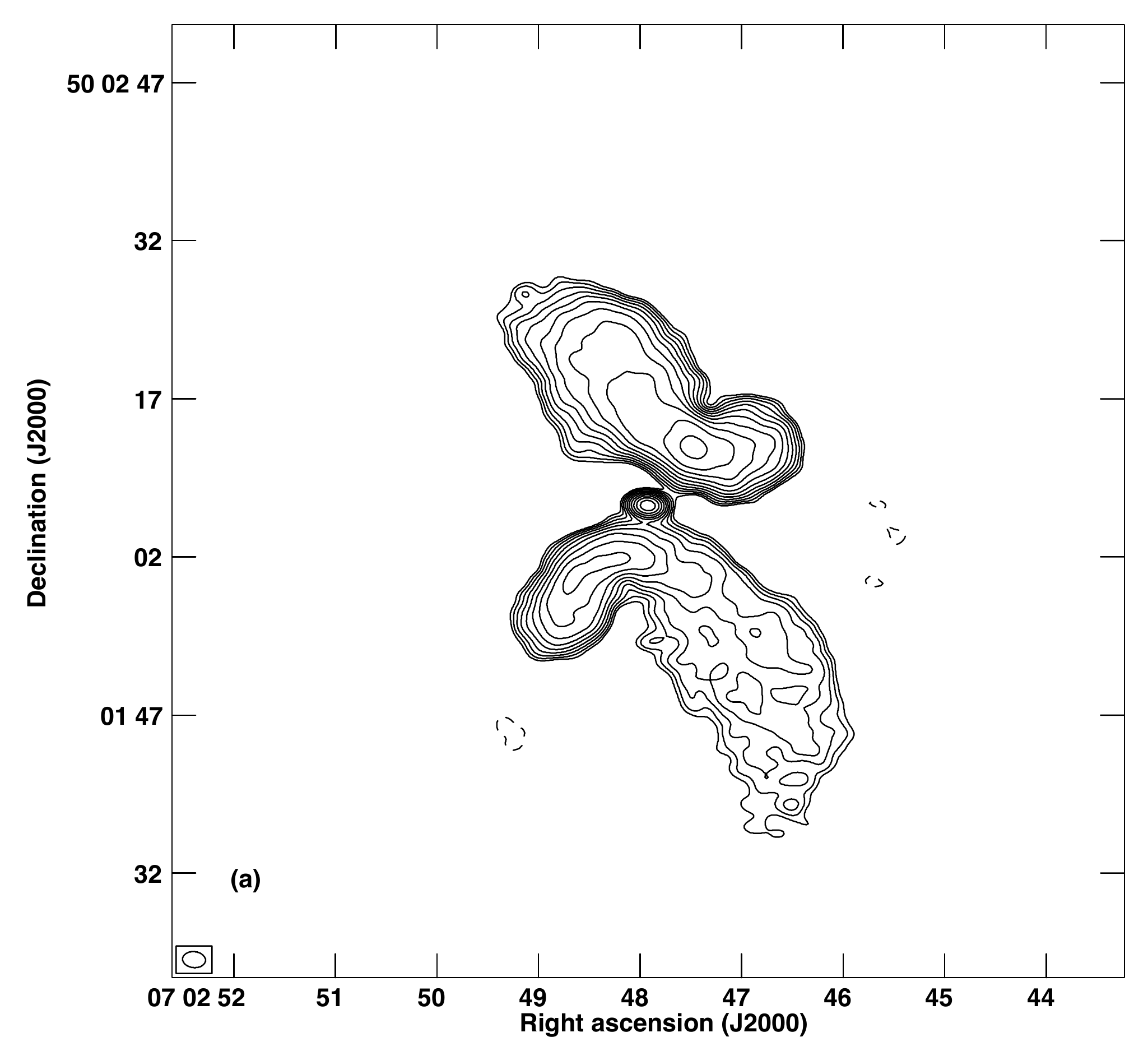}\\
\includegraphics[width=0.90\columnwidth]{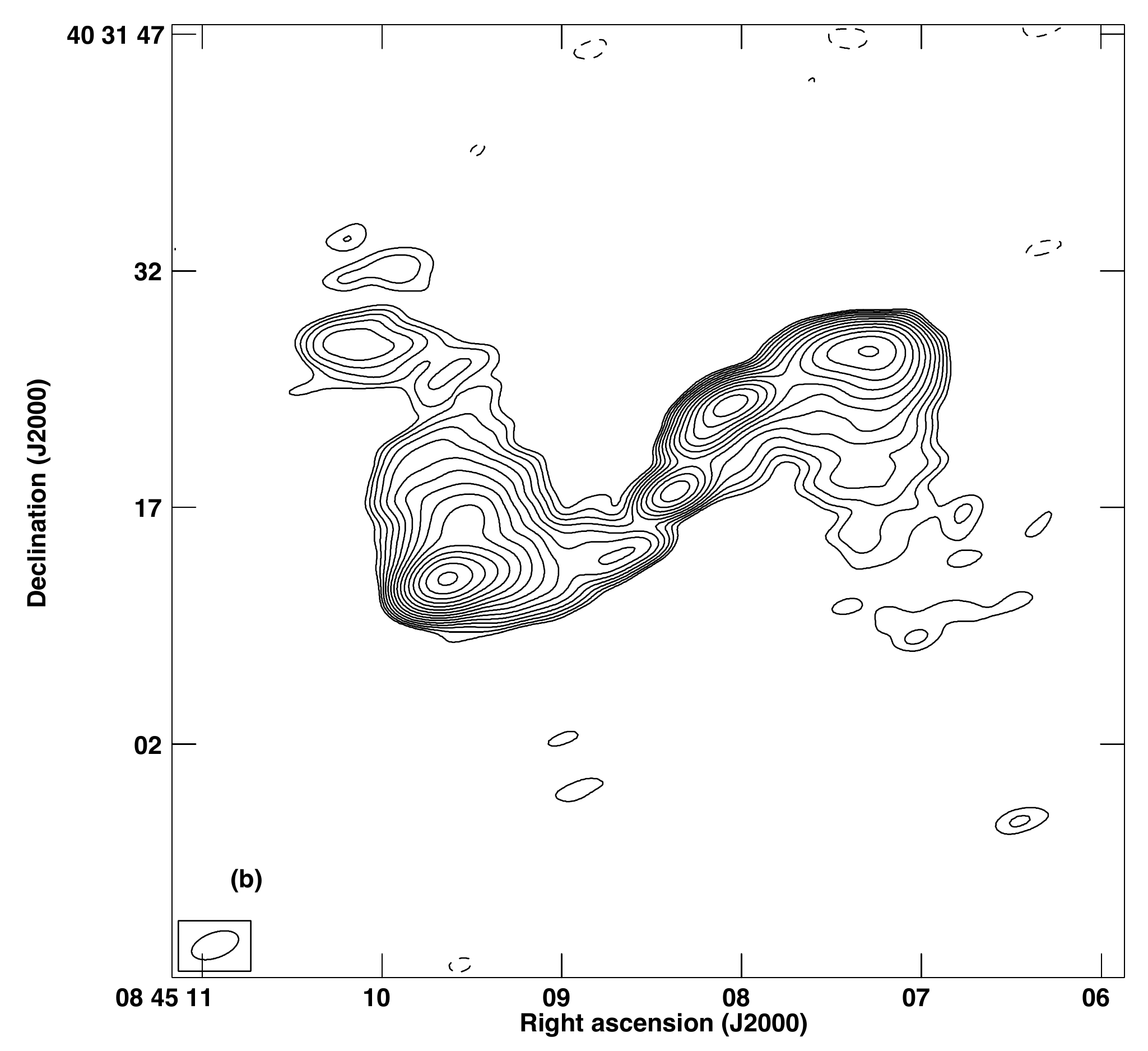}
\caption[J0702+5002]{ VLA L-band images of the radio galaxies (a) J0702+5002 (lowest contour = 0.2~mJy/beam, peak = 7.17~mJy/beam), and (b) J0845+4031 (lowest contour = 0.15~mJy/beam, peak  = 20.8~mJy/beam).  These are examples of sources where the off-axis emission occurs (a) at the inner ends of the main lobes and (b) at the outer ends of the lobes. Note the prominent inversion symmetry in each source's structure. \label{fig:notX}}
\end{figure}

An example of each of these classes is shown in Figure~\ref{fig:notX}. The sources with distortions near the inner ends of the on-axis lobes, and in which the off-axis emission on the two sides are not collinear, appear to be the best candidates for back flow deflections by thermal halos of the host ellipticals (\citet{Le1984}; but see \cite{Go2011}). It is possible that such inversion symmetric structures (which would classify as XRGs in the literature if the off-axis emission is long enough) might arise from rapid swings or flips of the jet axis during a merger; however, in explaining the variety in source structures this would require models where in some cases nuclei along with the host galaxy are displaced a considerable distance during mergers.  Additionally, in most cases the timescales associated with the flip would require to be fine tuned so that on one hand it is short enough to produce distinct wings and lobes and long enough for the older lobes of edge-brightened type to have morphed into edge-darkened or FR-I type. Hydrodynamic forces associated with the thermal gas halo of the host elliptical would now have to be invoked to create the emission gap between the two lobe pairs.  Most importantly, wings are almost always along the host minor axis and the active main lobes are towards the major axis \citep{C2002,Sa2009}, also for a sub-sample of the present set of sources, \citet{PaperI}): this requires that all of these inversion symmetric XRGs would have to be formed as a result of mergers that flip the axis from the minor to major axis.

\begin{figure}[ht] 
\includegraphics[width=0.90\columnwidth]{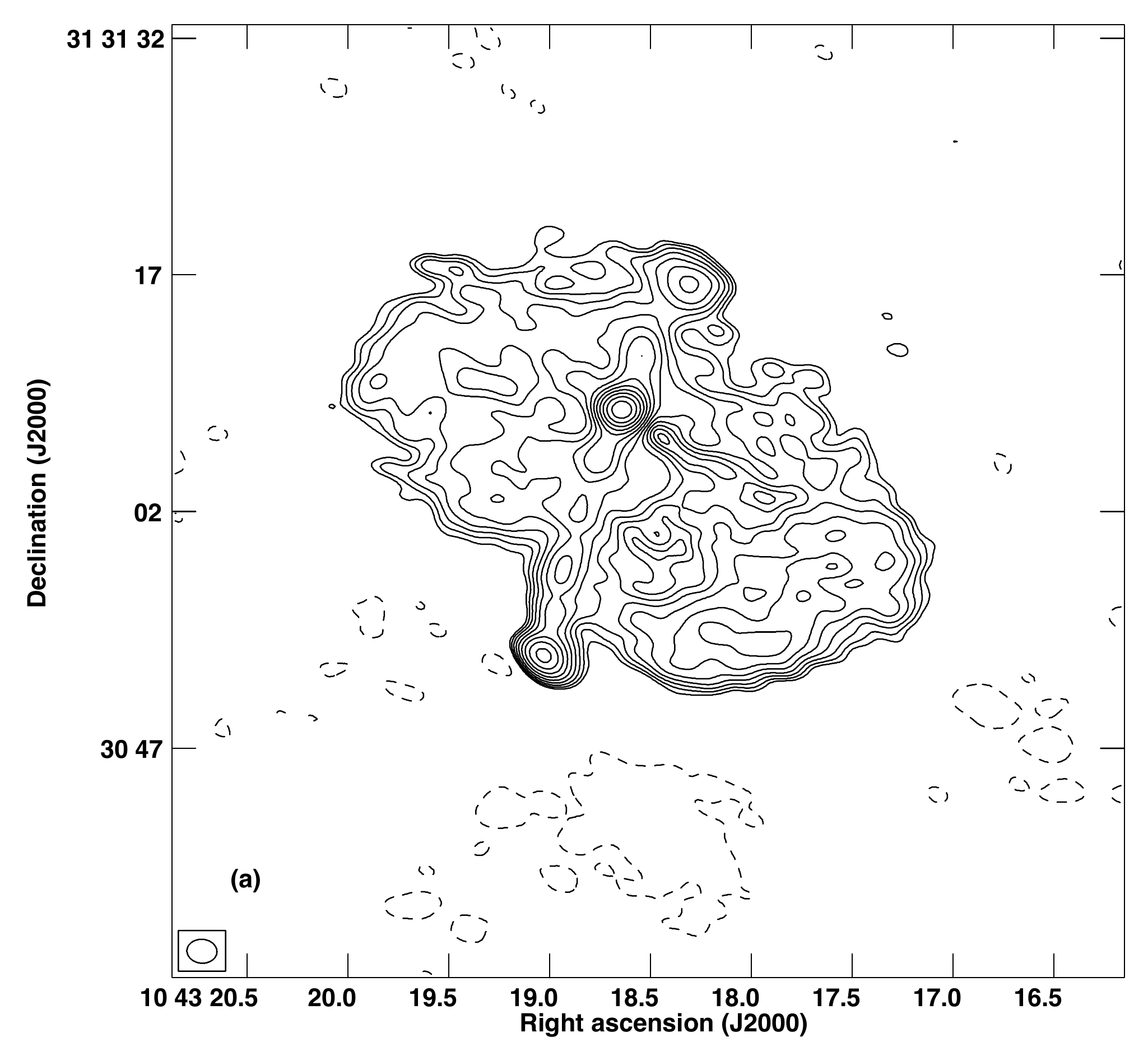}\\
\includegraphics[width=0.90\columnwidth]{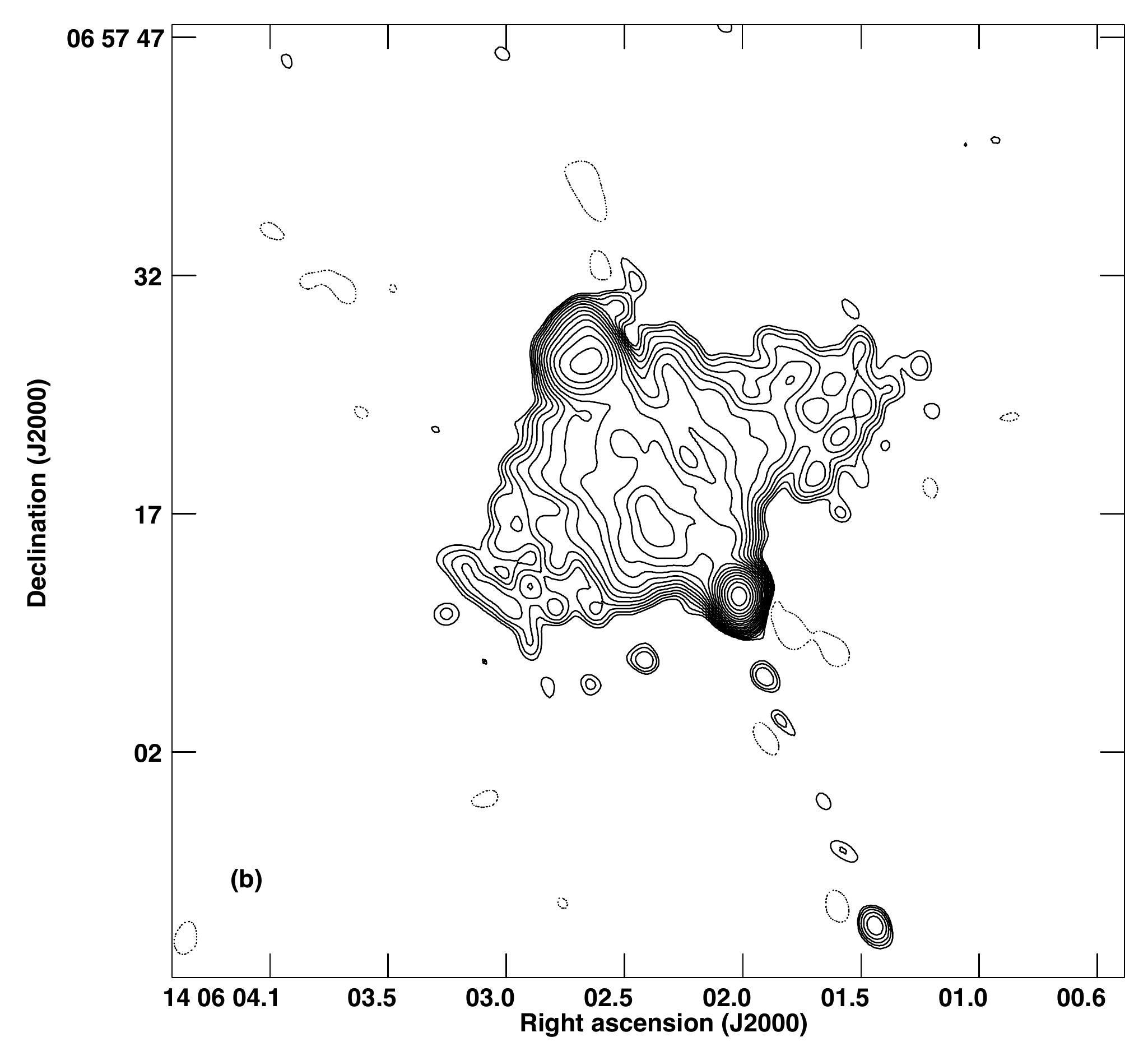}
\caption[J1043+3131]{ VLA L band images of radio galaxies that are candidates for genuine X-shape sources created by axis reorientations. (a) The upper panel shows J1043+3131 (lowest contour = 0.3~mJy/beam, peak  = 44.6~mJy/beam); (b) the lower panel shows J1406+0657 (lowest contour = 0.15~mJy/beam, peak  = 71.4~mJy/beam). 
\label{fig:genuineX}}
\end{figure}

On the other hand, the sources with distortions near the outer ends of the on-axis lobes appear to be candidates for cases of ``axis precession or drift,'' with radio structures that may signal the presence of binary black holes \citep{Be1980, D2014}, and thus be precursors of genuine X systems.  Excluding these two source categories, eleven sources (21\%) have been identified as genuine X-shaped radio galaxy candidates.\footnote{These sources are: J0144$-$0830, J1008+0030, J1015+5944, J1043+3131, J1327$-$0203, J1345+5233, J1406+0657, J1408+0225, J1606+0000, J1614+2817 and J1625+2705.} Two examples of candidate ``genuine X'' sources are shown in Figure~\ref{fig:genuineX}. 

We point out that our source morphology classification into outer-end and inner-end deviation sources can have implications for the general population of sources presently classified as XRGs. In our classification scheme several of the well-known XRGs will be classified as inner-end deviation sources and as argued above are more easily explained via the backflow deflection phenomenon rather than the spin-flip model. For the sources in Figure~\ref{fig:genuineX} the observations that we have do not indicate that the central transverse emission originates in either of the two lobes. Hence they are classified as X-shaped radio sources.

In the present paper we discuss the implications of these results for the GWB.

\section{GRAVITATIONAL RADIATION FROM SMBH COALESCENCE}

\subsection{Electromagnetic Signatures of Axis Reorientations}

Our imaging of a sample of low-axial ratio radio sources has allowed us to identify examples where the central transverse emission may be genuinely independent radio emission unrelated to the current activity. These are sources in which, within the limitations of sensitivity and resolution of the observations, the central transverse emission on the two sides cannot be traced to either of the two lobes, and appears as a swathe of emission through the central core or galaxy but at a large angle with respect to the main radio axis. We have identified 11 such sources; these sources are out of the sample of 52 sources whose radio structures we have examined carefully. These 52 are out of the sample of 100 sources that have been identified by \citet{C2007} to have low axial ratios and hence represent sources with substantial off axis emission.  The sample itself was selected from the FIRST survey that was searched for extended sources by selecting those with fitted major axis exceeding 15\arcsec~ (the FIRST beam was about 5\arcsec~FWHM); this yielded a total of 1648 extended radio sources.  The search of the FIRST images was limited to sources with peak flux density exceeding 5~mJy~beam$^{-1}$. Altogether this leads to an estimate that at most 1.3\% of extended radio sources might be genuine X-shaped radio sources. If in these sources (which we call genuine XRG's) the central swathe of off-axis emission arises because of axis reorientation alone then our observations point to an occurrence rate of less than 1.3\% for axis reorientations amongst the extended radio source population.

It may be noted that this rate of axis-reorientations in radio galaxies is much smaller than the occurrence rate (of 7\%) estimated by \citet{L1992} for XRGs (in the luminosity range of $0.3$--$30 \times 10^{25}$~W~Hz$^{-1}$).  However, Leahy \& Parma derived only the abundance fraction of radio galaxies with wings, without examining whether they were genuine X-shaped radio sources and without excluding sources where the off-axis emissions were obviously inversion-symmetric extensions of the lobes of a double lobed structure.  Based on the FIRST survey, the number of such candidate X-shaped sources was deemed to be 100 \citep{C2007}, which is indeed a 6\% fraction of the 1648 extended radio sources that were selected for examination.

On a related note, if the eight sources that appear to have off-axis emission connected to the outer lobe ends are representative of axis precession or drifts then our observations point to an occurrence rate of 0.9\% for axis drifts (occurring over a smaller angle of few tens of degrees).

It has been argued that sources with substantial off-axis emission that is traceable to the individual inner-lobes might form via a precession of the axis \citep{Go2011}.  While it is possible that precession may explain some source structures, it may be noted that \citet{Sa2009} showed that the radio structures in a more inclusive sample of X-shaped sources, which also included these sources with wings traceable to the inner-lobe ends as well, have a statistical relationship with the shape of the host elliptical in that the current radio axis is aligned closer to the major axis of the galaxy.  We see no obvious reason why axis reorientations ought to be almost always from an initial direction along the host minor axis to one along the major axis.  It is more reasonable therefore to suppose that the off-axis emission in most sources in such an inclusive sample arises from backflow deflections, as concluded by \citet{Sa2009}, and that the number of genuine X-shaped sources is a small subset of these candidates, as we suggest here.

\subsection{Gravitational Wave Background from SMBH Coalescence}

Prediction of the level of the GWB arising from the mergers of massive galaxies depends on estimation of the merger rate \citep{J2003}. This has been done in a variety of ways including studies of close pairs of galaxies \citep{C2000,W2009,X2012,J2014,K2014}, simulations of structure formation  \citep{K2008, Gu2011, RG2015}. For example, recent observations of a large number of close galaxy pairs \citep{K2014} show that a typical $L^*$ galaxy has undergone $\sim0.2-0.8$ mergers since $z \simeq 1$. Taking the mean of these numbers implies a major merger rate somewhat less than $0.1\mbox{ Gyr}^{-1}\mbox{ galaxy}^{-1}$. Estimation of the level of the GWB also depends on understanding the process of SMBH coalescence as driven by dynamical friction \citep{B1987} and gravitational radiation \citep{T1987}, as well as knowing how the last parsec problem is overcome \citep{Mi2003, C2014, V2014}. It would be useful to have an independent observational normalization for all of this, and the X-shaped radio galaxies offer one avenue in that direction.

\subsection{Constraints Arising from the XRG Population}

In this model, the coalescence rate for luminous ellipticals might be estimated by examining the occurrence rate of genuine X-shaped radio sources in the population of extended radio sources. If we use our finding above that at most 1.3\% of extended sources might have genuine X-shapes, then the occurrence rate of coalescence in their hosts is given by 0.013/$\tau$, where $\tau$ is the lifetime of relic lobes of radio galaxies.  If we adopt the value of $\tau = 10^{8}$~yr for the lifetimes of the relics (as assumed by \citet{M2002}) then we arrive at an estimate of 0.13~Gyr$^{-1}$galaxy$^{-1}$ as an upper limit to the event rate for radio galaxies, and presumably therefore for the parent population of massive ellipticals.\footnote{The radiation generated by SMBH coalescence in the few largest galaxies is expected to dominate the GWB \citep{Se2004}.} On the other hand, if we include the eight sources with off-axis emission traced to outer ends of lobes as sources that have undergone precession due to the presence of binary black holes \citep{D2014} then we estimate a coalescence rate of at most twice this value, of $\sim$ 0.2~Gyr$^{-1}$galaxy$^{-1}$. In either case the event rate is substantially smaller than the event rate of about 1~Gyr$^{-1}$galaxy$^{-1}$ usually assumed in calculations of the GWB from galaxy formation \citep{M2002}.

Our revision of the event rate is owing to the morphology classification adopted. If we consider the four examples of XRGs in \cite{M2002} in their Figure 2, only 3C\,223.1 would have been included as a potential ``genuine XRG'' as per our scheme. For the other three XRGs (3C\,52, 3C\,403, \& NGC\,326) given the available images we would note them as cases where the wings link with individual lobes rather than forming an independent transverse emission feature.

\section{SUMMARY AND CONCLUSIONS}

In \cite{PaperI} we analyzed 1.4  and 5~GHz archival VLA continuum data on a sample of 52 FIRST radio sources selected on the basis of low-axial ratio radio structures. In the current paper we discuss the impactions of the results for the gravitational wave background. Our principal results are:

\begin{enumerate} 

\item We argue that a large fraction of the sample (63\%) consists of sources where the off-axis emission is traced to individual lobes, which are more likely to be due to back flows or axis drifts rather than axis flips.

\item In this case the number of genuine X-shaped radio sources is at most eleven (21\%).

\item We then derive a five times lower occurrence rate for genuine X-shaped radio galaxies than the value estimated by \citet{L1992}.

\item If X-shaped radio galaxies arise because of major mergers alone then our observations point to an occurrence rate of less than 1.3\% for such mergers amongst the extended radio source population.

\item The low occurrence rate for major mergers suggests a significantly smaller event rate for coalescences of supermassive black holes of less than 0.13~Gyr$^{-1}$galaxy$^{-1}$ for hosts of radio galaxies. Inclusion of radio galaxies with off-axis emission traced to outer ends of lobes as sources that have undergone axis precession due to a close binary black hole only increases the coalescence rate by a factor of two.

\item If the class of sources with inner deviations are also cases of axis flips, then the event rate would be substantially greater, and similar to that derived by Merritt and Ekers (2002).  Indeed a measurement of the GWB might be a clue to the origins of the XRG structures.

\item Our estimate of the coalescence rate of giant ellipticals could be used to normalize calculations of the GWB from such events.

\end{enumerate}
 
 \section{FUTURE INVESTIGATIONS}

The elucidation of the structures in non-classical radio sources is important in that it has the power to yield event rates for SMBH coalescence and occurrence rates for a variety of phenomena that perturb the stability of the BH spin axis and interrupt the accretion flow that powers the beams.  The work presented in \cite{PaperI} and herein using primarily VLA archival data has been impactful in its conclusions for the event rate that determines the GWB, and clearly the conclusions could be substantially firmer with better data. Improved imaging in radio and optical broad band continuum, followed by derivation of source orientations in 3D with respect to the host ellipticity using depolarization distributions, and deriving the abundance statistics for the different types of off-axis distortions, may reveal the formation mechanisms for small-axial-ratio radio sources and hence provide a useful constraint on GWB from BH coalescence that accompanies interactions that reorient the spins of radio loud AGNs. 

While the improved resolution (compared to the FIRST survey) in the observations presented in \cite{PaperI} has helped in unraveling basic structure in several of the sources, the sensitivity is too limited in tracing fully the larger lobe and wing structures, all of which are critical in isolating genuine X-shaped radio galaxies. Deeper observations are needed to determine the structures more accurately than what have been imaged with the limited sensitivity archival observations presented here. Radio observations at multiple wavelengths would be useful in detecting gradients in spectral index within the lobes and hence indicate the time evolution in the structure.  We anticipate that with improved data the number of genuine candidates for axis reorientations would be further constrained, thereby reducing the event rates further and the discordance with formation simulations would be on firmer ground.

Critical to the estimation of the SMBH binary formation rate is the assumed lifetime of the relic radiation in an XRG. The estimate of $10^8\mbox{ yr}$ \citep{M2002} was based primarily on modeling of the spectral aging of the off-axis wings. With improved observations of the \cite{C2007} sample we will be able to make similar estimates for a larger sample of XRGs and thus firm up this critical parameter in the computation of the coalescence rate. Thus new JVLA observations are needed to test different formation models, estimate occurrence rate of axis reorientations, and hence also to place more accurate limits on the expected GWB signal. 

If models for axis flips are the explanations for inner lobe distortions, then we may explore the consequence of the inference that flips almost always occur from minor to major axis.   Indeed, giant radio sources do have their axes along the minor axis of the hosts \citep{Sa2009}. This leads to the suggestion that the minor axis is the stable state for SMBH spins.   Explorations of the torques and tidal couplings between the large-scale structure of the host elliptical and the inner accretions disk and hence BH spin axis may suggest an explanation for this phenomenology.

\section{ACKNOWLEDGMENTS}

The National Radio Astronomy Observatory is a facility of the National Science Foundation, operated under cooperative agreement by Associated Universities, Inc.  D.\ H.\ R.\ gratefully acknowledges the support of the William R. Kenan, Jr.\ Charitable Trust. We thank the referee, Prof.\ David Merritt, for his useful suggestions.


\end{document}